# Extremely high Q-factor toroidal metamaterials

**Alexey A. Basharin[1]**, Vitaliy Chuguevskiy[2], Nikita Volsky[1], Maria Kafesaki[3], Eleftherios N. Economou[3], Alexey V. Ustinov[1,4]

[1] National University of Science and Technology MISiS, Leninskiy prosp. 4, Moscow, 119049, Russia

[2] Voronezh state technical University, Universitetskaya pl. 1, Voronezh, 394036, Russia

[3] Foundation for Research and Technology Hellas (FORTH), and University of Crete, Heraklion, Crete, Greece

[4] Physikalisches Institut, Karlsruhe Institute of Technology, 76131 Karlsruhe, Germany

e-mail of the presenter: alexey.basharin@gmail.com

*Abstract* – **We demonstrate that, owing to the unique topology of the toroidal dipolar mode, its electric/magnetic field can be spatially confined within subwavelength, externally accessible regions of the metamolecules, which makes the toroidal planar metamaterials a viable platform for high Q-factor resonators due to interfering toroidal and other dipolar modes in metamolecules.**

## I. INTRODUCTION

In this paper, we present experimental and numerical results on complementary electric- and magnetic-type planar metamaterials composed of metamolecules possessing toroidal dipole moments. Electromagnetic field within toroidal metamolecules remains spatially confined to subwavelength scale which minimizes radiation losses. Toroidal metamaterials are viable platform for sensing, enhancement of light absorption and optical nonlinearities [1]. We envision that addition of the toroidal ingredients may improve coherence of superconducting qubits and quantum metamaterials [2].

Planar toroidal metamaterials are promising due to simple fabrication in comparison with the classical 3D toroidal inclusions. However, a planar topology has a major drawback associated with the dimensional limitation. Indeed, pure toroidal moment is unexcitable in 2D planar inclusions. Thus, the overall response is accompanied by a significant contribution of the magnetic quadrupole and electric octupole [3] moments. In this paper, we for the first time demonstrate theoretically and experimentally that extremely high Q-factors can be achieved in the planar pseudo-toroidal metamaterials due to significant concentration of electric and magnetic fields. Interestingly, these properties are linked to the interference between electric dipole, toroidal and magnetic quadrupole moments (Fig. 1a,b) and between toroidal and magnetic quadrupole moments (Fig. 2a,b), respectively.

## II. "ELECTRIC"-TYPE TOROIDAL METAMATERIAL

Here we study two types of metamaterials. The first one is composed of "electric" metamolecules (Fig. 1a).The electric metamolecules are planar conductive structures consisting of two symmetric split loops. The incident plane wave excites circular currents **j** along the loops lead to a circulating magnetic moment and, as a result, to a toroidal moment **T**. Two side gaps also support a magnetic quadrupole moment **Qm**. Moreover, due to the central gap electric moment **P** can be exited in metamolecule (Fig. 1d). Interestingly, it was concluded in [3] that isolated toroidal dipole moment is unachievable in planar geometry and magnetic quadrupole moment **Qm** always accompanies metamaterial response. We agree with this conclusion. At the same time, destructive/constructive interference between **T**, **Qm** and **P** gives us unique effect as very strong **E**- field localization inside central gap (Fig. 1b). We note that this allows us to achieve extremely high Q-factors of such type of metamaterial at the frequency of simultaneous excitation of these moments (Fig. 1c).

The theoretical and numerical investigation of the proposed electric-type toroidal metamaterial was done in microwave regime. The experimental sample was fabricated from a slab of steel by laser cutting method, self-supporting without any dielectric substrate. The metamolecules have diameter d= 15 mm and are placed tightly close to each other. The central gap is 1.2 mm, the side gaps are 2 mm. At the frequency close to 10 GHz we observe extremely narrow resonance of S21 (transmission), both in experiment and numerical simulations (Fig 1c). The resonance is associated with a very high electric field concentrated in the central gap (Fig 1b) partly due to the toroidal response of the system (Fig 1d). Moreover, multipole expansion results presented in Fig 1d show that, besides the existence of the toroidal dipole **T**, strong electric **P** and magnetic quadrupole moments **Qm** are excited and effectively interfere with the toroidal dipole **T**. We stress that the internal *Q*-factor of the resonance dip (measured at 3 dB level above the minimum) is ~ 40000 at room temperature.

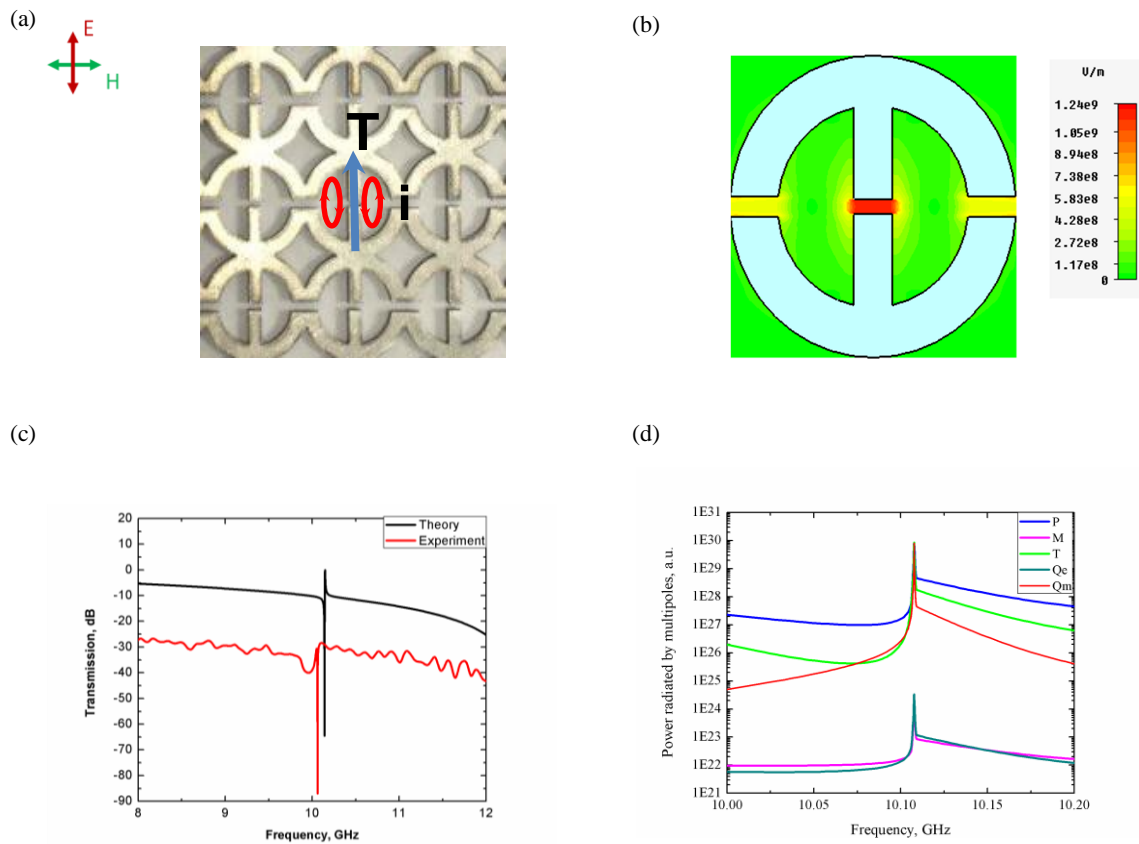

Fig. 1. (a)- A fragment of an "electric" type planar metamaterial sample supporting toroidal dipolar excitation. Red arrows show displacement currents **j** induced by the vertically polarized plane wave, blue arrow shows Toroidal dipole moments **T** of the metamolecule. (b)- The absolute value of the Electric field **E**. (c)- Transmission S21, dB spectra spectra calculated for the metamaterial (black curve) and obtained from experiment, (red curve). (d) Contributions of the five strongest multipolar excitations of the metamaterial to the power radiated by the array.

### III. "Magnetic"-type toroidal meta materials

The second studied structure investigated we call "magnetic" type metamolecule (Fig. 2a). It is the inverted and rotated Babinet-variant of the first structure (Fig 1a). In contrast to the first case, here we expect very strong localization of magnetic field instead electric field. The magnetic field lines are whirling around the central junction of the metamolecule (Fig. 2b) due to interference between toroidal **T** and magnetic quadrupole moment **Qm** (Fig. 2d). Importantly, this configuration allows us to reduce electric moment **P**. Hence, we observe very strong magnetic field localization (Fig. 2b). This localization is associated with a very narrow and pronounced transmission dip, as is shown in Fig. 2c where the corresponding transmission response, both theoretical and experimental, is demonstrated. We are not aware of similar results reported elsewhere. Thus, very high magnetic field localization in subwavelength area is reported here for the first time and is due to toroidal response. The measured internal *Q*-factor is ~ 57000 at room temperature.

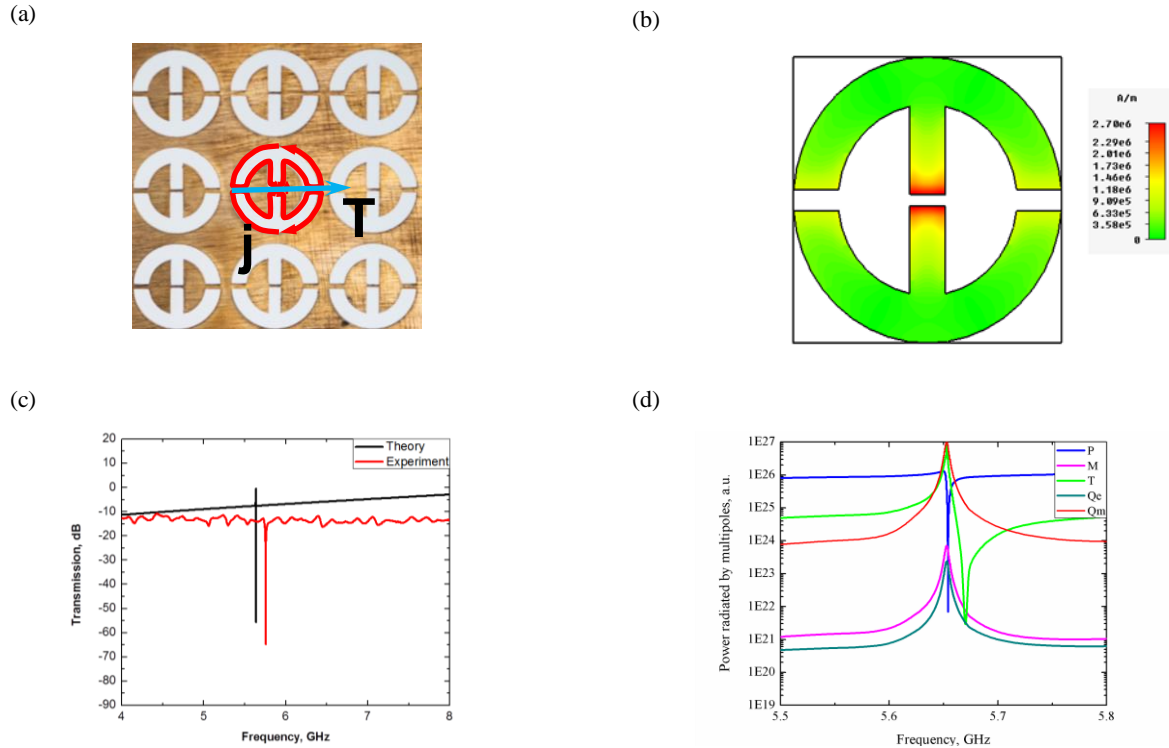

Fig. 2. (a)- A fragment of a "magnetic" type planar metamaterial sample supporting toroidal dipolar excitation. Red arrows show displacement currents **j** induced by the vertically polarized plane wave, blue arrow shows toroidal dipole moments **T** of the metamolecule. (b)- The absolute value of the magnetic field **H**. (c)- Transmission S21, dB spectra calculated for the metamaterial (black curve) and obtained experimentally, (red curve). (d) Contributions of the five strongest multipolar excitations of the metamaterial to the power radiated by the array.

## IV. CONCLUSION

In summary, here we studied planar toroidal metamaterials in the microwave frequency range. Both electric and magnetic metamaterials demonstrate extremely high Q-factors due to strong concentrated electric/magnetic fields induced through the toroidal response and interference with other multipoles. We stress that proposed metamaterials are promising for strong nonlinear excitations and thus can be made tunable.

## ACKNOWLEDGEMENT

The authors gratefully acknowledge the financial support of the Ministry of Education and Science of the Russian Federation in the framework of Increase Competitiveness Program of NUST «MISiS» (№ K4-2015-031), the Greek General Secretariat for Research and Technology in the framework of the project ERC-02: EXEL-6260, the European Union in the framework of the project GA-320081 PHOTOMETA, the EU-Russian project EXODIAGNOS, as well as the Deutsche Forschungsgemeinschaft (DFG).

## REFERENCES

[1] N. Papasimakis, V. A. Fedotov, V. Savinov, T. A. Raybould & N. I. Zheludev, Electromagnetic toroidal excitations in matter and free space, Nature Materials 15, 263–271 (2016)
[2] P. Macha, G. Oelsner, J.-M. Reiner, M. Marthaler, S. Andre, G. Schoen, U. Huebner, H.-G. Meyer, E. Ilichev and A. V. Ustinov, Implementation of a quantum metamaterial using superconducting qubits, Nature Commun. 5, 5146 (2014)
[3] Savinov, V. Novel toroidal and superconducting metamaterials PhD thesis, Univ. Southampton (2014).